\documentclass[aps,prx,twocolumn,showpacs,superscriptaddress]{revtex4-2}
\usepackage{graphicx}
\usepackage{amsmath}
\usepackage{amssymb}
\usepackage{colordvi}
\usepackage{mathrsfs}
\usepackage{bm}
\usepackage{verbatim}
\usepackage{dcolumn}
\usepackage{bm}
\usepackage{epsfig}
\usepackage{subfigure}
\usepackage{dsfont}
\usepackage{bbding}
\usepackage[colorlinks,linkcolor=blue,anchorcolor=blue,citecolor=blue]{hyperref}

\begin{document}
	\title{Pronounced in-plane anomalous Hall effect with vanishing out-of-plane response in $\bf{Cr_{1.2}Te_{2}}$}  
\author{Wenzhi Peng}
\affiliation{International Center for Quantum Design of Functional Materials, School of Emerging Technology, University of Science and Technology of China, Hefei, Anhui 230026, China}
\author{Zheng Liu}
\affiliation{Department of Physics, University of Science and Technology of China, Hefei, Anhui 230026, China}
\affiliation{CAS Key Laboratory of Strongly-Coupled Quantum Matter Physics, University of Science and Technology of China, Hefei, Anhui 230026, China}
\author{ShaSha Wang}
\affiliation{Hefei National Research Center for Physical Sciences at the Microscale, Department of Materials Science and Engineering, University of Science and Technology of China, Hefei, Anhui 230026, China}	
\author{Haolin Pan}
\affiliation{International Center for Quantum Design of Functional Materials, School of Emerging Technology, University of Science and Technology of China, Hefei, Anhui 230026, China}	
\author{Changlong Wang}
\affiliation{Hefei National Research Center for Physical Sciences at the Microscale, Department of Materials Science and Engineering, University of Science and Technology of China, Hefei, Anhui 230026, China}	
\author{Xiangbiao Shi}
\affiliation{Hefei National Research Center for Physical Sciences at the Microscale, Department of Materials Science and Engineering, University of Science and Technology of China, Hefei, Anhui 230026, China}
\author{Jiahao Han}
\affiliation{Center for Science and Innovation in Spintronics, Tohoku University, Sendai, Japan}
\author{Qian Niu}
\affiliation{Department of Physics, University of Science and Technology of China, Hefei, Anhui 230026, China}
\affiliation{CAS Key Laboratory of Strongly-Coupled Quantum Matter Physics, University of Science and Technology of China, Hefei, Anhui 230026, China}	
\author{Yang Gao}
\affiliation{International Center for Quantum Design of Functional Materials, School of Emerging Technology, University of Science and Technology of China, Hefei, Anhui 230026, China}	
\affiliation{Department of Physics, University of Science and Technology of China, Hefei, Anhui 230026, China}
\affiliation{CAS Key Laboratory of Strongly-Coupled Quantum Matter Physics, University of Science and Technology of China, Hefei, Anhui 230026, China}
\author{Bin Xiang}
\email[Correspondence author:~~]{binxiang@ustc.edu.cn}
\affiliation{Hefei National Research Center for Physical Sciences at the Microscale, Department of Materials Science and Engineering, University of Science and Technology of China, Hefei, Anhui 230026, China}
\author{Dazhi Hou}
\email[Correspondence author:~~]{dazhi@ustc.edu.cn}
\affiliation{International Center for Quantum Design of Functional Materials, School of Emerging Technology, University of Science and Technology of China, Hefei, Anhui 230026, China}			
\affiliation{Department of Physics, University of Science and Technology of China, Hefei, Anhui 230026, China}	

	
	\begin{abstract}
We report an unconventional anomalous Hall regime in the van der Waals ferromagnet $\rm{Cr_{1.2}Te_{2}}$, in which the anomalous Hall effect (AHE) is present for in-plane magnetization but absent for out-of-plane magnetization. In this purely in-plane regime, the anomalous Hall signal exhibits a threefold angular dependence during both in-plane and out-of-plane rotations of the magnetization, which cannot be accounted for by the conventional dipolar contribution but instead requires an octupolar contribution. Although the octupolar term qualitatively captures the observed behavior, the experimentally extracted octupole differs quantitatively from first-principles calculations based solely on the intrinsic Berry-curvature mechanism, indicating an essential role for extrinsic scattering processes.

	\end{abstract}
	
\maketitle

The anomalous Hall effect (AHE) is a fundamental transport phenomenon in magnetic materials driven by spin--orbit coupling (SOC), which originates from either the intrinsic Berry curvature in momentum space or the extrinsic contributions from skew scattering and side-jump processes~\cite{Hall1880,Nagaosa2010,Dheer1967,Jungwirth2003,Tian2009,Nakatsuji2015,Liu2018,Luttinger1958,Berger1970,Yao2004,Chen2014}. Regardless of the underlying mechanism, the AHE in ferromagnets is conventionally associated with a finite out-of-plane magnetization component and thus exhibits a sinusoidal angular dependence on the magnetization direction~\cite{Nagaosa2010,Zeng2006,Wang2020NatComm,Guillet2021,Chen2025}. Recent experiments, however, have revealed that an anomalous Hall signal can also be generated by a purely in-plane magnetization---the so-called in-plane anomalous Hall effect (IPAHE)~\cite{Friedland2001,Xiaosongwu2024_PRL,Nishihaya2025,Peng2024Octupole,Chen2025Fe211,M.Uchida_2024_PRL,Kao2026}. Such behavior has been reported and systematically investigated in a variety of ferromagnetic systems~\cite{Friedland2001,Xiaosongwu2024_PRL,Nishihaya2025,Peng2024Octupole,Chen2025Fe211,M.Uchida_2024_PRL}, and more recently observed in a topological semimetal heterostructure with reduced crystal symmetry~\cite{Kao2026}. A salient feature common to all these observations is the coexistence of IPAHE and OPAHE, with the IPAHE invariably accompanied by a finite OPAHE~\cite{Friedland2001,Xiaosongwu2024_PRL,Nishihaya2025,Peng2024Octupole,Chen2025Fe211,M.Uchida_2024_PRL,Kao2026}. 

On the theoretical side, the description of the AHE in ferromagnets had been focused on the dipolar contribution, under which the anomalous Hall conductivity is proportional to the out-of-plane magnetization component and therefore vanishes at in-plane magnetization, leaving the origin of the IPAHE unexplained~\cite{Hall1880,Nagaosa2010,Dheer1967,Jungwirth2003,Tian2009,Nakatsuji2015,Liu2018,Chen2025,Roman2009}. Important progress was made by identifying Weyl-point splitting as the microscopic origin of the IPAHE in $\mathrm{Fe_3Sn_2}$ and $\mathrm{EuCd_2Sb_2}$~\cite{Xiaosongwu2024_PRL,M.Uchida_2024_PRL}, offering valuable insight into how band topology gives rise to an in-plane Hall response. However, extending this picture to systems without nontrivial band topology, such as Fe where the IPAHE has also been observed~\cite{Friedland2001,Peng2024Octupole,Chen2025Fe211}, called for a more general framework. Liu \textit{et al.} developed a symmetry-based theory that reformulates the anomalous Hall conductivity as a multipolar expansion in terms of the magnetization direction, consisting of a leading dipolar term and higher-order multipolar contributions~\cite{Liu2025,Liu2025b}. Within this framework, a reduced crystal symmetry that permits dipolar anisotropy can give rise to the IPAHE, as recently observed in the topological semimetal heterostructure~\cite{Kao2026}. In systems with higher crystal symmetry, such as $\mathrm{Fe_3Sn_2}$, $\mathrm{SrRuO_3}$(111), and Fe~\cite{Friedland2001,Xiaosongwu2024_PRL,Nishihaya2025,Peng2024Octupole,Chen2025Fe211,M.Uchida_2024_PRL}, the dipolar term vanishes for in-plane magnetization, and the IPAHE instead arises solely from the multipolar terms. Since the multipolar contributions emerge only at higher orders in SOC and are therefore expected to remain much smaller than the dipolar term, the IPAHE should always coexist with a finite and even larger OPAHE~\cite{Friedland2001,Xiaosongwu2024_PRL,Nishihaya2025,Peng2024Octupole,Chen2025Fe211,M.Uchida_2024_PRL,Kao2026,Liu2025}.

\begin{figure}
	\includegraphics[width=8cm,angle=0]{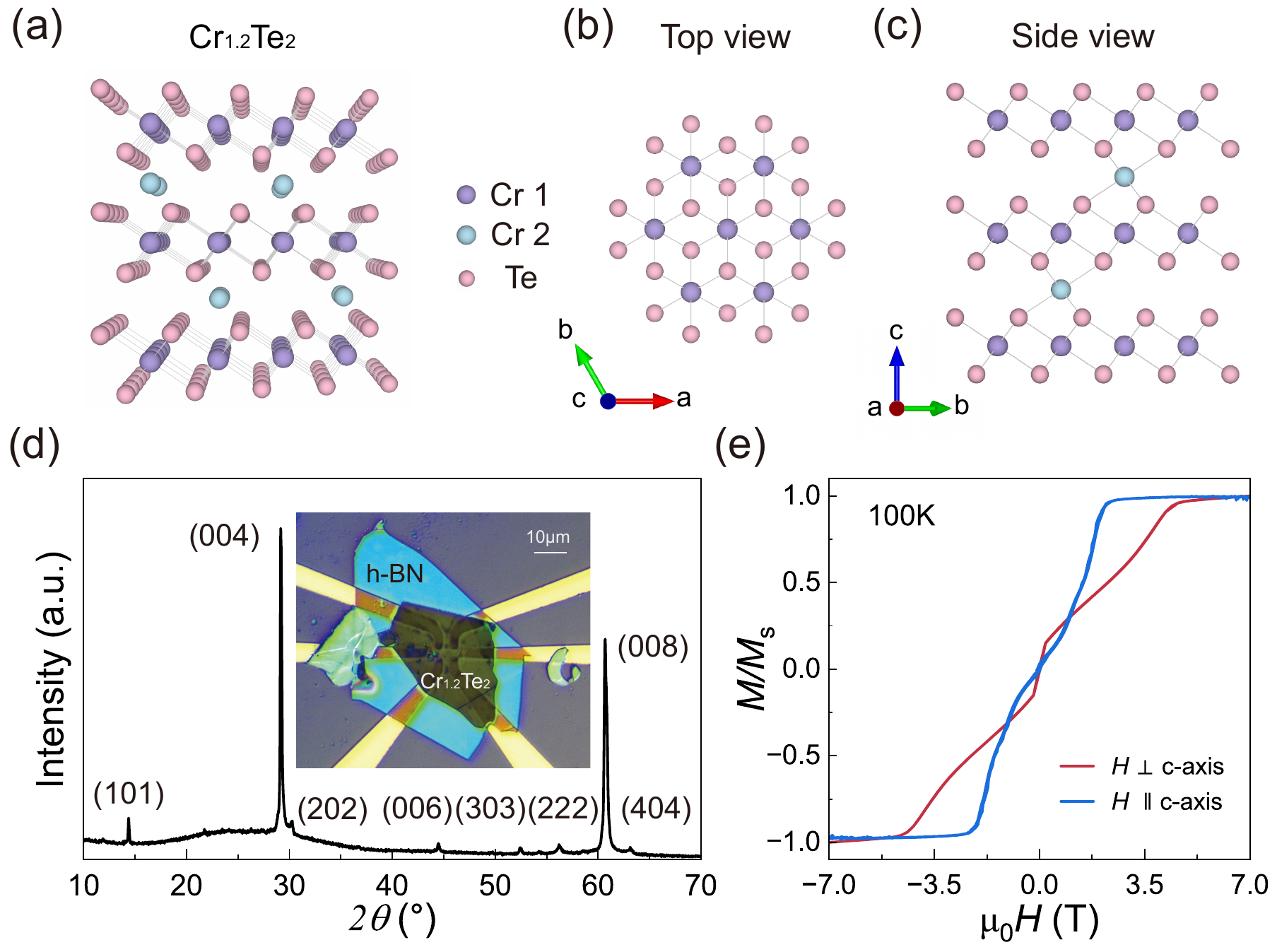}
	\caption{(a) Schematic illustration of the crystal structure of $\rm{Cr_{1.2}Te_{2}}$, where the intercalated Cr atoms with fractional occupancy are shown as blue spheres. Atomic structure of $\rm{Cr_{1.2}Te_{2}}$ viewed from the top (b) and side (c) directions\cite{Huang2021a,Huang2021b}.
		(d) Powder X-ray diffraction results for $\rm{Cr_{1.2}Te_{2}}$. Inset: Optical image of a 45-nm-thick $\rm{Cr_{1.2}Te_{2}}$ Hall bar device encapsulated with h-BN.
		(e) The $\bm {M-H}$ curves of $\rm{Cr_{1.2}Te_{2}}$ bulk sample at 100 K, with $\bm H$ applied parallel and perpendicular to the $\rm c$ axis. 
	}
	\label{Fig.1}
	
\end{figure}

Here we report a striking violation of this expectation. In the van der Waals ferromagnet $\mathrm{Cr_{1.2}Te_2}$\cite{Sun2020,Zhang2021CrTe2,Meng2021,Huang2021a,Huang2021b}, the conventional OPAHE vanishes at 25~K while a pronounced IPAHE persists, yielding a divergent IPAHE-to-OPAHE ratio in sharp contrast to previous observations~\cite{Friedland2001,Xiaosongwu2024_PRL,Nishihaya2025,Peng2024Octupole,Chen2025Fe211,M.Uchida_2024_PRL,Kao2026}. By lowering the temperature, the OPAHE decreases and reverses the sign around 25~K, whereas the IPAHE increases monotonically. At the crossover temperature where the OPAHE vanishes, the anomalous Hall response becomes entirely governed by the threefold-symmetric IPAHE. To elucidate the underlying mechanism, we decompose the measured anomalous Hall conductivity into dipolar and multipolar contributions, revealing that the dipolar coefficient decreases and reverses sign upon cooling, whereas the multipolar coefficients remain finite and grow monotonically. Notably, despite this anomalous hierarchy among the coefficients, the multipolar anisotropy of the AHE successfully captures the observed threefold angular dependence under both in-plane and out-of-plane magnetization rotations, confirming the validity of the symmetry-based framework in describing the magnetization-direction dependence of the AHE. However, first-principles calculations based on the intrinsic Berry-curvature mechanism yield a dipolar contribution much larger than the higher-order multipolar terms, preserving the conventional hierarchy and thus failing to account for the anomalous dominance of the IPAHE observed in this work. This discrepancy suggests that extrinsic scattering plays an essential and previously underappreciated role in governing the hierarchy between the IPAHE and the OPAHE.

Figure~1a shows the crystal structure of $\mathrm{Cr_{1.2}Te_2}$ (space group $P\bar{3}m1$\cite{Huang2021a,Huang2021b}), with top and side views in FIGs. 1b and 1c revealing a clear threefold rotational symmetry about the c-axis. Figure 1d presents the X-ray diffraction (XRD) pattern confirming high crystalline quality, with the inset showing an optical image of the Hall-bar device fabricated by dry transfer. Figure~1e shows the magnetization hysteresis loop ($\bm M$--$\bm H$) measured at 100~K, demonstrating the ferromagnetic nature of $\mathrm{Cr_{1.2}Te_2}$. At lower temperatures, the saturation field increases and the magnetic signal becomes noisier (FIG.~S1).

\begin{figure}
	\includegraphics[width=8.5cm,angle=0]{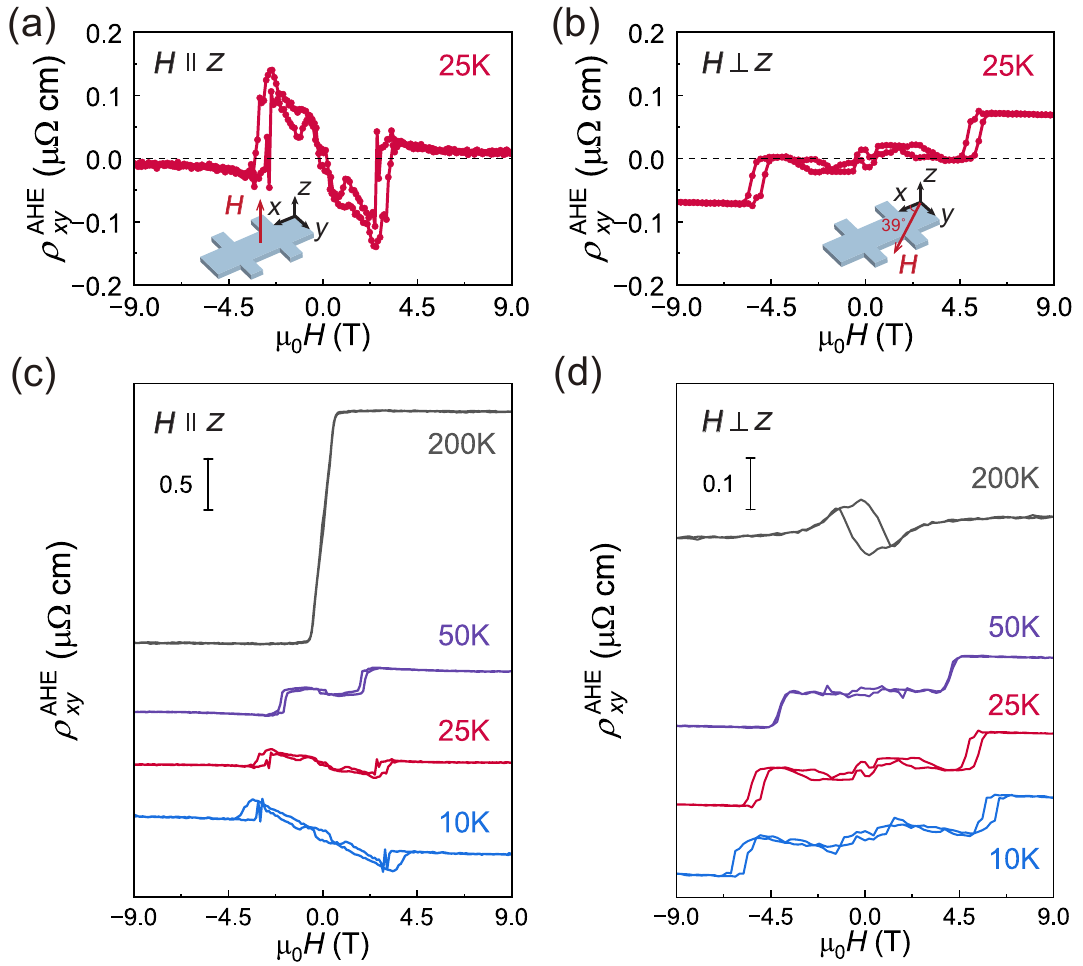}
	\caption{(a),(b) Magnetic-field dependence of $\rho^{\rm AHE}_{xy}$ measured at 25 K with $\bm H \parallel z$ (a) and $\bm H \perp z$ (b) for $\lvert \bm H \rvert \leq 9$ T. Insets illustrate the measurement geometries.
		(c),(d) Temperature evolution of $\rho^{\rm AHE}_{xy}$ measured with $\bm H \parallel z$ (c) and $\bm H \perp z$ (d) from 200 K to 10 K.
	}		
	\label{Fig.2}
\end{figure}

Figures 2a and 2b show the magnetic-field dependence of the anomalous Hall resistivity $\rho^{\rm AHE}_{xy}$ at 25 K for $\bm H\parallel z$ and $\bm H\perp z$, respectively. For the out-of-plane configuration (FIG. 2a), $\rho^{\rm AHE}_{xy}$ nearly vanishes for  $|\bm H|>$ 4.5 T in the magnetization saturation regime. In contrast, a pronounced IPAHE is observed for the in-plane configuration as in FIG. 2b. The ratio of the IPAHE-to-OPAHE reaches $\sim10$, far exceeding previously reported values (typically smaller than 1)~\cite{Xiaosongwu2024_PRL,Nishihaya2025,Peng2024Octupole,Chen2025Fe211,M.Uchida_2024_PRL}. Figures 2c and 2d show the OPAHE and IPAHE signal measured at different temperatures, respectively. As the temperature decreases, the OPAHE amplitude progressively decreases and reverses sign at lower temperatures, whereas the IPAHE increases monotonically. The $\rho^{\rm AHE}_{xy}$ is obtained by subtracting the linear normal Hall background via high-field linear fitting (FIG. S2a). The distinct temperature dependences suggest different underlying mechanisms for the OPAHE and IPAHE.

In the dipolar framework of the AHE, the anomalous Hall conductivity linearly scales with the magnetization, $\sigma_i^{\rm AHE}=p_{ij}m_j$, with $i,j= x,y,z$. In $\rm{Cr_{1.2}Te_{2}}$, the $C_{3z}$ symmetry imposes strong constraints on the dipolar tensor $p_{ij}$, restricting it to a single nonzero component $p_{zz}$~\cite{Peng2024Octupole}. Consequently, the dipolar contribution reduces to $\sigma_{z}^{\rm AHE}=p_{zz}m_{z}$, which cannot account for the pronounced IPAHE observed at $\bm H\perp z$. To determine the symmetry-allowed form of the anomalous Hall response beyond the dipolar framework, we consider the constraints imposed by the $C_{3v}$ point group symmetry of $\rm{Cr_{1.2}Te_{2}}$. Within the multipolar expansion of the anomalous Hall conductivity in magnetization space, the Onsager reciprocal relation dictates that all the odd-order multipolar terms contribute to the anomalous Hall response\cite{Onsager1931,Onsager1931b}. Retaining terms up to third order, the anomalous Hall conductivity can be written as:	
	\begin{align}\label{eq:SigmaC3v}
		\sigma_{z}^{\rm AHE}
		&=(p_{zz}-\frac{3}{2}o_{zzzz})m_{z}+\frac{5}{2}o_{zzzz}m_{z}^3 \notag\\
		&+o_{zyyy}m_{y}^3-3o_{zyyy}m_{x}^2m_{y}
	\end{align}
Here $p_{zz}$ denotes the dipolar coefficient, while $o_{zzzz}$ and $o_{zyyy}$ represent the octupolar contributions. According to Eq.~\ref{eq:SigmaC3v}, the AHE is governed by both $p_{zz}$ and $o_{zzzz}$ with out-of-plane magnetization, whereas it arises solely from the $o_{zyyy}$ term for in-plane magnetization. As a result, the OPAHE and IPAHE are controlled by different coefficients, hinting at the origin of their distinct temperature dependences observed in FIGs.~2c and 2d. In particular, the disappearance of the OPAHE could be understood as a consequence of the competition between  $p_{zz}$ and $o_{zzzz}$, while the IPAHE remains finite due to its different origin. 

Eq.~\ref{eq:SigmaC3v} can be rewritten in terms of the magnetization direction angles $(\theta,\varphi)$ defined in the insets of the FIGs. 3a and 3d as:	
	\begin{align}\label{eq:C3vAngular}
		\sigma_{z}^{\rm AHE}
		&=(p_{zz}+\frac{3}{8}o_{zzzz})\cos\theta
		+\frac{5}{8}o_{zzzz}\cos3\theta \notag\\
		&+o_{zyyy}\Big(\frac{3}{4}\sin\theta+\frac{1}{4}\sin3\theta\Big)\sin 3\varphi
	\end{align}	
which can be directly compared with the AHE measurement of magnetization rotation. For out-of-plane rotation, the onefold component $\cos\theta$ contains contributions from both $p_{zz}$ and $o_{zzzz}$. However, the octupolar term introduces an additional threefold component $\cos3\theta$, which enables its separation from the dipolar contribution, but has not been observed in previous measurements. For in-plane rotation, the response reduces to a threefold dependence $\propto \sin 3\varphi$, originating solely from $o_{zyyy}$. Therefore, although dipolar and octupolar contributions are blended in the anomalous Hall response, their distinct angular dependences enable a quantitative decomposition. In particular, the emergence of threefold angular components in both in-plane and out-of-plane configurations provides a unified signature of the octupolar contribution to the AHE.

\begin{figure}
	\includegraphics[width=8.0cm,angle=0]{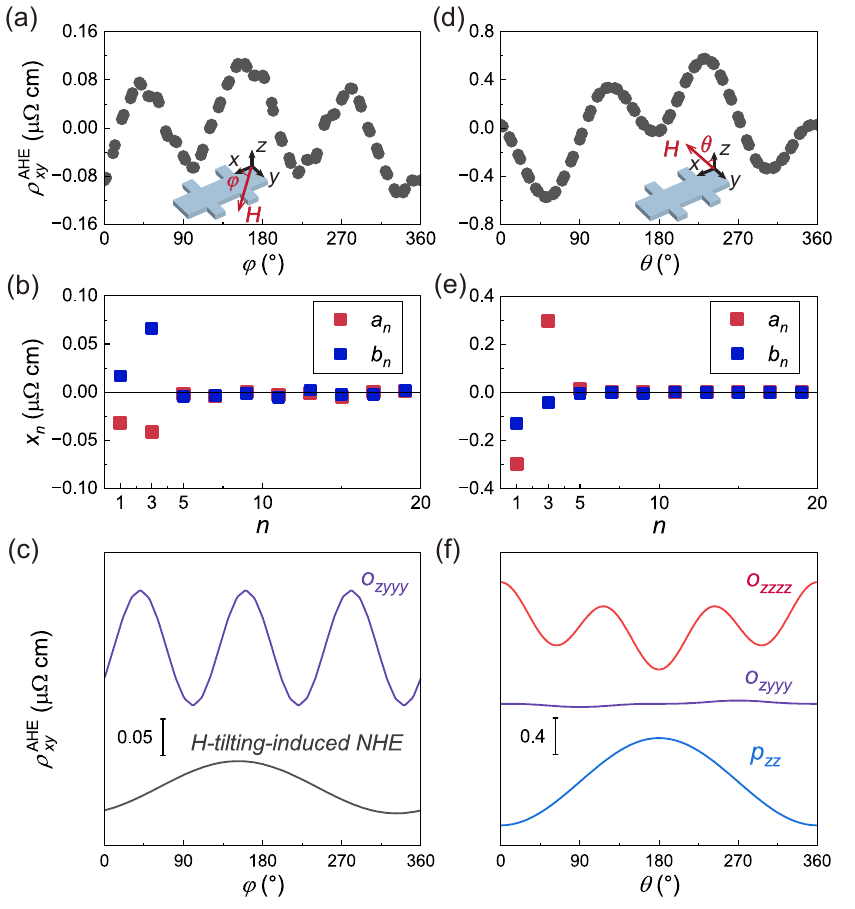}
	\caption{(a), (d) Field angular dependence of $\rho^{\rm AHE}_{xy}$, measured as $\bm H$ rotates in the $x-y$ and $x-z$ planes at 25 K with $\lvert \bm H \rvert = 6.5$ T (a) or $\lvert \bm H \rvert = 9$ T (d); the corresponding measurement geometries are illustrated in the insets.
		(b), (e) Harmonic amplitudes extracted from the Fourier expansion $f(\varphi)=\sum_{ n}[a_{n}\cos(n\varphi)+b_{n}\sin(n\varphi)]$ of the angular-dependent data in (a) and (d).
		(c), (f) Decomposition of $\rho^{\rm AHE}_{xy}$ into symmetry-allowed contributions according to Eq.~\ref{eq:C3vAngular}. In (c), the black curve denotes the out-of-plane noise arising from an unintentional $\bm H$-tilting out of the Hall plane, while the purple curve corresponds to the octupolar contribution $o_{zyyy}$. In (f), the blue curve denotes the dipolar contribution $p_{zz}$, whereas the red and purple curves represent the octupolar contributions $o_{zzzz}$ and $o_{zyyy}$, respectively.}		
	\label{Fig.3}
\end{figure}

Figure 3a shows the angular dependence of $\rho^{\rm AHE}_{xy}$ under in-plane rotation at 25 K, exhibiting a pronounced threefold symmetry, in qualitative agreement with the octupolar contribution discussed above. 
The data are analyzed using a Fourier expansion $f(\varphi)=\sum_{ n}[a_{n}\cos(n\varphi)+b_{n}\sin(n\varphi)]$, with harmonic amplitudes summarized in FIG. 3b. A dominant $n=3$ component is clearly resolved, consistent with the expected $\sin3(\varphi+\varphi_0)$ dependence from the $o_{zyyy}$ term, where $\varphi_0\approx -9^\circ$ reflects a small misalignment between the current direction and the crystallographic $\rm a-$axis. Higher-order harmonics ($n\geq 5$) remain within the noise level, indicating that contributions beyond the octupolar term are below the detection threshold\cite{Peng2024Octupole}. A finite $n=1$ component is also observed, originating from an unintentional $\bm H$-tilting out of the plane that induces a normal Hall effect (NHE)\cite{Xiaosongwu2024_PRL,Peng2024Octupole,Eden2019}. Accordingly, the signal in FIG. 3a can be decomposed into a dominant contribution from the octupolar term $o_{zyyy}$ and a minor $\bm H$-tilting-induced background, as shown in FIG. 3c. As confirmed in FIG. S2c, this $\bm H$-tilting-induced contribution is negligibly small compared to the NHE for $\bm H \parallel z$, corresponding to a field misalignment of ~0.87$^\circ$.

For out-of-plane rotation in the $x-z$ plane, the normal Hall effect introduces a sinusoidal background due to the out-of-plane component of the magnetic field, which is subtracted prior to analysis (FIG. S2b). The resulting angular dependence of $\rho^{\rm AHE}_{xy}$ is shown in FIG. 3d. Remarkably, in addition to the familiar $\sin\theta$ dependence~\cite{Nagaosa2010,Wang2020NatComm,Guillet2021,Chen2025}, a pronounced threefold component is clearly resolved, in agreement with Eq.~\ref{eq:C3vAngular}. This feature is further confirmed by the Fourier expansion in FIG. 3e, representing the first experimental observation of a threefold angular dependence in the OPAHE. According to Eq.~\ref{eq:C3vAngular}, although both $p_{zz}$ and $o_{zzzz}$ contribute to the onefold component, only the octupolar term can generate a threefold component. This distinct angular dependence provides a direct handle to separate the dipolar and octupolar contributions. We accordingly decompose the anomalous Hall response into multipolar contributions, as shown in FIG. 3f. The contribution from $o_{zyyy}$ enters via the finite azimuthal offset $\varphi_0$, but remains much smaller and is negligible in the out-of-plane rotation.

\begin{figure}
	\includegraphics[width=8.5cm,angle=0]{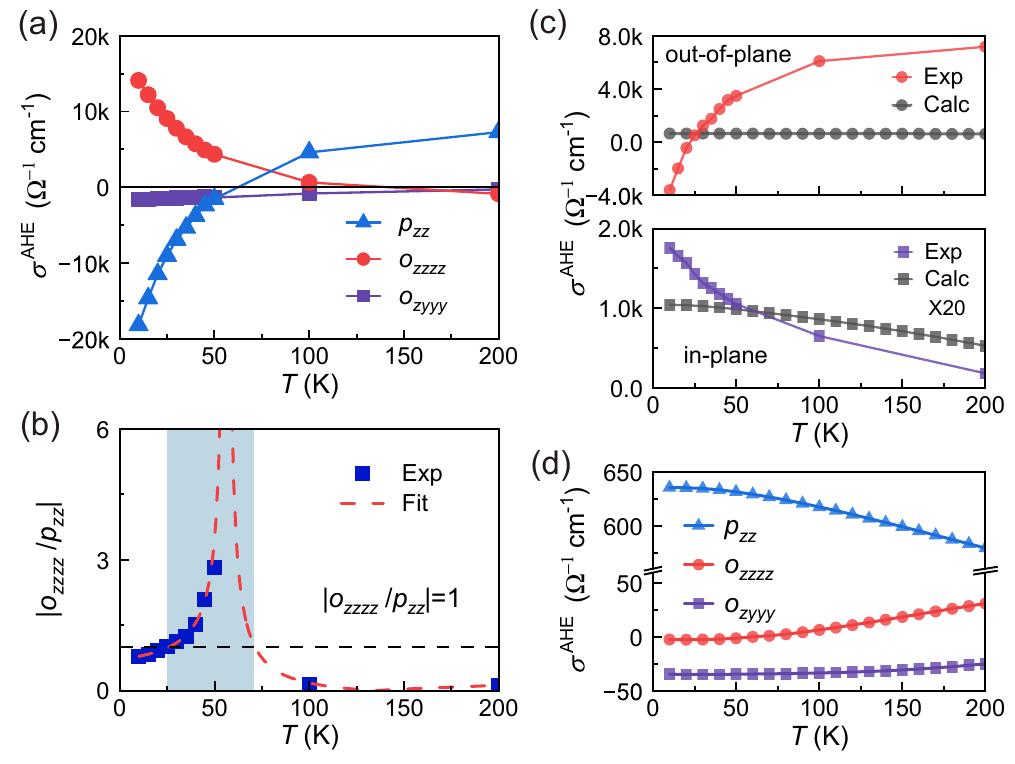}
	\caption{(a) Temperature dependencies of the dipolar ($p_{zz}$, blue) and octupolar ($o_{zyyy}$, purple; $o_{zzzz}$, red) coefficients extracted from the angular-dependent $\sigma^{\rm AHE}$. (b) Temperature dependence of the ratio $|o_{zzzz}/p_{zz}|$. The shaded region highlights the octupole-dominated regime. The dashed curve is calculated from the ratio of the guide-to-the-eye fits to $p_{zz}$ and $o_{zzzz}$ shown in (a). (c) Temperature dependences of $\sigma^{\mathrm{AHE}}$ for the out-of-plane (upper panel) and in-plane (lower panel) configurations, comparing experimental data with first-principles calculations. The calculated values in the lower panel are multiplied by 20 to facilitate comparison. (d) Temperature dependence of the multipolar coefficients calculated from first-principles.}

	\label{Fig.4}
\end{figure}

To gain further insight into the microscopic origin of the AHE, we examine the temperature dependence of the dipolar and octupolar coefficients extracted from angular-dependent measurements, as summarized in FIG. 4a. As the temperature decreases from 200 K, $p_{zz}$ is strongly suppressed and changes its sign around 50 K, reminiscent of behaviors reported in previous studies of ferromagnets where the OPAHE exhibits vanishing or sign reversal~\cite{Fang2003,Mathieu2004,Huang2012,Wu2013,Siddiquee2023,Huang2025}. In contrast, both $o_{zzzz}$ and $o_{zyyy}$ remain finite and do not exhibit a sign reversal below 100 K. The contributions of $p_{zz}$ and $o_{zzzz}$ to the anomalous Hall response cancel around 25~K, while $o_{zyyy}$ remains finite, leaving an AHE regime solely sensitive to the in-plane magnetization. This naturally explains the coexistence of a vanishing OPAHE and a pronounced IPAHE observed in this work. As shown in FIG.~4b, $|o_{zzzz}|/|p_{zz}|$ exceeds unity over a broad temperature range, marking the first experimentally observed octupole-dominated AHE. 

We next examine whether the intrinsic mechanism can account for the observed octupole-dominated AHE. Figure~\ref{Fig.4}c compares the temperature dependence of the calculated OPAHE and IPAHE based on the intrinsic Berry-curvature mechanism with the corresponding experimental results~\cite{SM}. For the OPAHE, the experimental $\sigma^{\mathrm{AHE}}$ decreases upon cooling and changes sign below 25~K, whereas the calculated value remains nearly temperature-independent and positive throughout. For the IPAHE, the calculated $\sigma^{\mathrm{AHE}}$ is nearly two orders of magnitude smaller than the experimental value. Figure~\ref{Fig.4}d shows the calculated multipolar coefficients, which preserve the perturbative hierarchy across the entire temperature range~\cite{Liu2025}: $p_{zz}$ remains dominant and positive, while $o_{zzzz}$ and $o_{zyyy}$ stay substantially smaller, all falling nearly two orders of magnitude below the experimental values shown in Fig.~\ref{Fig.4}a. Clearly, the intrinsic Berry-curvature mechanism fails to provide even a qualitative description of the AHE observed here. These findings indicate that extrinsic scattering processes play an essential and previously underappreciated role in governing the multipolar anisotropy of the AHE, the microscopic origin of which remains an important open question for future theoretical work. Furthermore, establishing an anomalous Hall scaling law capable of disentangling intrinsic and extrinsic contributions in the octupole-dominated AHE poses a significant experimental challenge~\cite{Tian2009,Hou2015}.


\begin{acknowledgments}
This work was supported by the National Natural Science Foundation of China (Grant Nos. 12234017, 12074366, and 12374164). D. Hou and Y. Gao were supported by the Fundamental Research Funds for the Central Universities (Grant Nos. WK9990000116 and WK2340000102). Z. Liu was supported by the National Natural Science Foundation of China (Grant Nos. 11974327 and 12004369), the Fundamental Research Funds for the Central Universities (Grant Nos. WK3510000010 and WK2030020032), the Anhui Initiative in Quantum Information Technologies (Grant No. AHY170000), and the Innovation Program for Quantum Science and Technology (Grant No. 2021ZD0302800). The sample fabrication was supported by the USTC Center for Micro- and Nanoscale Research and Fabrication. The authors thank K. J. Dai and L. F. Wang for the magnetization measurement.
\end{acknowledgments}

\end{document}